\documentclass[aps,prl,fleqn,twocolumn,superscriptaddress,floatfix]{revtex4}

\usepackage{epsfig}
\usepackage{amsmath,amssymb,amsfonts}

\begin{document}

\title{Possible Resolutions of the $D$-Paradox}

\author{C. Nonaka}
\affiliation{Department of Physics, Duke University, Durham, NC 27708}

\author{B.~M\"uller}
\affiliation{Department of Physics, Duke University, Durham, NC 27708}

\author{S.~A.~Bass}
\affiliation{Department of Physics, Duke University, Durham, NC 27708}
\affiliation{RIKEN BNL Research Center, Brookhaven National Laboratory, 
             Upton, NY 11973}

\author{M.~Asakawa}
\affiliation{Department of Physics, Osaka University,
             Toyonaka 560-0043, Japan}

\date{\today}

\begin{abstract}
We propose possible ways of explaining the net charge
event-by-event fluctuations
in Au+Au collisions at the Relativistic Heavy Ion Collider within a 
quark recombination model. We discuss various methods of estimating
the number of quarks at recombination and their implications for the
predicted net charge fluctuations. We also discuss the possibility of
diquark and quark-antiquark clustering above the deconfinement temperature. 
\end{abstract}


\maketitle

Fluctuations of the net electric charge of all particles emitted
into a specified rapidity window have been proposed as a possible
signal for the formation of deconfined quark matter in relativistic
heavy ion collisions \cite{Asakawa:2000wh,Jeon:2000wg}. The argument
at the basis of this proposal is that charge fluctuations in a
quark-gluon plasma are expected to be significantly smaller (by
a factor $3-4$) than in a hadronic gas. Because the net charge 
contained in a given volume is locally conserved and can only be 
changed by particle diffusion, thermal fluctuations generated within 
the deconfined phase could survive hadronization and final state
interactions. Quantitative estimates of the diffusion of net charge
showed that the survival of these fluctuations from an early stage
of the collision requires a moderately large rapidity window 
\cite{Shuryak:2000pd}. 

The most widely used measure for the entropy normalized net charge
fluctuations is the $D$ measure \cite{Jeon:2000wg}:
\begin{equation}
D = 4 \langle (\Delta Q)^2 \rangle / N_{\rm ch},
\label{Deq}
\end{equation}
where $\langle (\Delta Q)^2 \rangle$ denotes the event-by-event 
net charge fluctuation within a given rapidity window $\Delta y$, 
and $N_{\rm ch}$ is the total number of charged particles emitted
in this window. For a free plasma of quarks and gluons
$D\approx 1$, while for a free pion gas $D\approx 4$.
For the comparison with experimental data a number of corrections 
for acceptance and global charge conservation
must be applied to the expression for $D$ \cite{Bleicher:2000ek}.
The relation of the $D$-measure to other measures of net charge
fluctuations has been discussed by various authors
\cite{Mrowczynski:2001mm,Zaranek:2001di,Nystrand:2003ed}.

Several experiments have measured net charge fluctuations in heavy
ion collisions at the CERN Super-Proton Synchrotron (SPS) and at
the Relativistic Heavy Ion Collider (RHIC) in Brookhaven
\cite{STAR_fluc,PHENIX_fluc,Sako:2004pw,Alt:2004ir}. The
results for $D$ are generally somewhat smaller than 4, but much 
larger than the value predicted for a free quark-gluon gas.
For example, the STAR collaboration has measured $D=2.8\pm 0.05$
in central Au+Au collisions at $\sqrt{s_{NN}}=$ 130 GeV 
\cite{STAR_fluc}, before applying corrections for global charge 
conservation and other effects \cite{Bleicher:2000ek}.
The PHENIX experiment measured net charge fluctuations in a
limited azimuthal acceptance window around midrapidity, which
extrapolate to a value $D\approx 3$ \cite{PHENIX_fluc}.
These results are surprising, because many other observables
indicate that a deconfined quark-gluon plasma is formed in these
collisions. 

Bia{\l}as has argued that the measured values of $D$ could
be compatible with the net charge fluctuations in a deconfined quark
phase, if hadronization proceeds according to simple valence
quark counting rules \cite{Bialas:2002hm} and if gluons do not
play an active role in the hadronization. Indeed, hadron abundances
measured in relativistic heavy ion collisions at the SPS and
RHIC are well described by combinatorial quark recombination
models, such as ALCOR \cite{Biro:1994mp}. We here pursue 
this idea further and explore various scenarios of valence quark 
recombination in order to better understand how the puzzle posed
by the measured value of $D$ can be resolved. We also discuss the
constraints on such a resolution from the measured final-state
entropy and the second law of thermodynamics.

The recombination of thermalized valence quarks has recently been 
proposed as the dominant mechanism for the production of hadrons 
with transverse momenta of a few GeV/$c$ in Au+Au collisions at RHIC 
\cite{Vo02,Fr03,Gr03,Hw03,Mo03,Ko04}. The RHIC data have provided 
compelling evidence for this hadronization mechanism 
\cite{PHENIX_reco,STAR_reco}. 
Valence quark recombination explains the enhancement of baryon emission, 
compared with meson emission, in the range of intermediate transverse 
momenta (roughly from 2 to 5 GeV/$c$), and it naturally describes
the observed hadron species dependence of the elliptic flow in the 
same momentum region in terms of a universal elliptic flow curve for 
the constituent quarks \cite{Fr03}.

When one wants to describe hadronization by quark recombination not
only at intermediate momenta, but over the entire hadron momentum
range, except for very large momenta where parton fragmentation is 
thought to dominate, entropy becomes an important consideration. 
The naive application of recombination can easily entail a violation
of the second law of thermodynamics, because the number of independent
particles decreases when valence quarks recombine into hadrons.
Greco {\em et al.} \cite{Gr03} have argued that this problem may 
be circumvented by including the decay of hadronic resonances, such 
as the $\rho$-meson, into the calculation of the entropy balance. 
We next discuss the entropy problem in a more comprehensive manner 
by considering the entropy content in realistic models of the hadronic 
phase (the resonance gas model) and the quark phase (lattice-QCD 
calculations).

The equilibrium entropy per particle is only a fixed constant (3.60 for 
bosons, 4.20 for fermions) for free massless particles. For particle 
with mass, the entropy per particle is a function of the particle 
mass $m$ and the temperature $T$. For $m/T > 3$ a good approximation is
$S/N = 3.50 + m/3T$. 
The inclusion of mass is important,  since a large fraction of the hadrons created
at the moment of hadronization is quite heavy -- the average hadron
mass at chemical freezeout is about 800 MeV/$c^2$ in the absence of medium
induced modifications of the hadron masses. The
average value of the entropy per hadron therefore significantly exceeds
the canonical value $(S/N)_0 \approx 4$ in thermal equilibrium. 
Including only the pseudoscalar and vector ground state meson nonets 
and the octet and decuplet baryons, one finds $\langle S/N \rangle = 
4.72$ at $T = T_c = 170$ MeV; including all known meson and baryon 
resonances, one obtains $\langle S/N \rangle = 5.15$. 

An estimate of the final-state entropy per unit rapidity produced 
in central $\sqrt{s_{NN}}=130$ GeV Au+Au collisions has recently been 
derived from experimental data (hadron yields, spectra, and source radii) 
by Pal and Pratt: $dS/dy = 4450 \pm 400$ \cite{P+P}. Using the measured 
charged multiplicity of $dN_{\rm ch}/dy = 526 \pm 2 {\rm (stat)} \pm 36 
{\rm (syst)}$ \cite{STAR_fluc}, this value can be converted into an
estimate of the final entropy per particle of $S/N \approx (dS/dy)/
(1.5 dN_{\rm ch}/dy) \approx 5.64 \pm 0.6$, which is slightly higher, but
agrees within error with the equilibrium value for the full resonance 
gas at chemical freezeout. Indeed, numerical simulations of the
reactions among hadrons in the hadron gas phase \cite{B+D} show
that the particle number increases only slightly (by about 5\%) due 
to interactions after hadronization, compared with the number obtained 
by letting all hadrons formed at chemical freezeout decouple and decay 
without reinteraction. A modest increase
in the value of $S/N$ during the hadron gas phase can easily be 
understood as the effect of volume expansion, since the entropy per
particle for fixed particle number increases logarithmically with
the volume: $S/N = S_{\rm eq}/N + \ln(V/V_{\rm eq})$.

At the same time, the entropy content of the quark phase is strongly 
reduced due to interactions near $T_c$. Recently, the CP-PACS 
collaboration \cite{CPPACS} and the Bielefeld group \cite{Bielefeld} 
have calculated the pressure and energy density at finite
temperature and zero chemical potential on the lattice. 
The Bielefeld simulation was done keeping the ratio $m_q/T$ fixed, 
while the one by the CP-PACS collaboration was performed with the ratio of     
pseudo-scalar to vector meson masses, $m_{\rm PS}/m_{\rm V}$ fixed.
The CP-PACS collaboration used temporal lattice sizes $N_t = 4, 6$ 
and $N_f = 2$ quark flavors; the Bielefeld group used only 
$N_t = 4$, but explored a range of flavor multiplicities 
$N_f = 2,~2+1,~{\rm and}~3$. In order to extract physical quantities 
from lattice QCD calculations, extrapolations (to the thermodynamic 
limit, continuum limit, etc.) are mandatory. Because of a finer lattice 
spacing at a given temperature, the CP-PACS simulation for $N_t = 6$ 
is closer to the continuum limit and may be slightly 
better suited for the purpose of extracting the entropy density near 
$T_c$. We use these results here. However, we should keep in mind 
that all lattice data for thermodynamic quantities are still obtained
with unphysically large quark masses.

We list the obtained values of $\varepsilon/T^4$, $P/T^4$,
$\varepsilon/\varepsilon_{\rm SB}$, $P/P_{\rm SB}$, and
$s/s_{\rm SB}$, at $m_{\rm PS}/m_{\rm V} = 0.65 $ and $0.80$ below in 
Tables \ref{cppacs-1} and \ref{cppacs-2}. $\varepsilon$, $P$, and 
$s$ denote the energy density, pressure, and entropy density,
respectively, and $\varepsilon_{\rm SB}$, $P_{\rm SB}$, and $s_{\rm SB}$,
are their values for the free gas of massless
quarks and gluons on the lattice used in the simulation.
The statistical errors are shown only for $s/s_{\rm SB}$.
Obviously, the size of the systematic errors are still substantial.
However, at the same time, it is obvious that the entropy density
of the quark-gluon plasma is considerably suppressed with respect 
to the corresponding Stefan-Boltzmann (SB) value near $T_c$.

\begin{table}[thb]
\caption{$m_{\rm PS}/m_{\rm V} = 0.65$ case. All calculations are performed
with $(N_t = 6)$. Statistical errors are shown only for $s/s_{\rm SB}$. }
\begin{center}
\begin{tabular}{crcccc}
\hline
\hline
$T/T_c$ & 
$\varepsilon/T^4$ & 
$P/T^4$ & 
$\varepsilon/\varepsilon_{\rm SB}$ &
$P/P_{\rm SB}$ &
$s/s_{\rm SB}$ \\
\hline
0.92 &  8.03 & 0.88 & 0.489 & 0.172 & $0.413 \pm 0.051$ \\
1.04 & 11.91 & 1.28 & 0.725 & 0.250 & $0.612 \pm 0.043$ \\
1.30 & 13.67 & 2.75 & 0.833 & 0.536 & $0.762 \pm 0.038$ \\
1.61 & 14.44 & 3.35 & 0.879 & 0.653 & $0.826 \pm 0.038$ \\
2.00 & 12.93 & 3.96 & 0.787 & 0.772 & $0.784 \pm 0.038$ \\
\hline
\hline
\end{tabular}
\end{center}
\label{cppacs-1}
\end{table}

\begin{table}[thb]
\caption{Same as Table \ref{cppacs-1} except $m_{\rm PS}/m_{\rm V} = 0.80$.}
\begin{center}
\begin{tabular}{crcccc}
\hline
\hline
$T/T_c$ & 
$\varepsilon/T^4$ & 
$P/T^4$ & 
$\varepsilon/\varepsilon_{\rm SB}$ &
$P/P_{\rm SB}$ &
$s/s_{\rm SB}$ \\
\hline
0.80 &  3.73 & 0.17 & 0.227 & 0.033 & $0.181 \pm 0.054$ \\
0.89 &  1.91 & 0.26 & 0.116 & 0.061 & $0.101 \pm 0.041$ \\
1.12 & 12.08 & 1.62 & 0.736 & 0.316 & $0.636 \pm 0.044$ \\
1.38 & 11.98 & 2.66 & 0.730 & 0.519 & $0.679 \pm 0.042$ \\
1.67 & 11.80 & 3.54 & 0.719 & 0.690 & $0.712 \pm 0.039$ \\
\hline
\hline
\end{tabular}
\end{center}
\label{cppacs-2}
\end{table}

From these results we conclude the following:
\begin{enumerate}
\item The entropy per particle in the hadronic gas, and therefore 
      the entropy content of the hadronic phase at chemical freezeout,
      is considerably larger than often assumed.
\item The entropy density of the quark phase is significantly 
      suppressed near $T_c$, most likely due to correlations among
      the quasiparticles caused by their strong interactions. 
\end{enumerate}
These two conclusions make the recombination picture of 
hadronization more compatible with the entropy constraint. Namely, 
if the quark-gluon plasma at hadronization consists of strongly
interacting quasi-particles (possibly constituent quarks) with strong
correlations, and if many of the hadrons created at hadronization 
are heavy, quark recombination and the concomitant particle number 
decrease can be more easily reconciled with
the second law of thermodynamics.

As we have seen, the increased  
entropy per hadron in the massive resonance gas may well allow for 
an isentropic transition from the deconfined to the confined phase 
by (sudden) quark recombination.
At present, however, we cannot directly 
compare the entropy content of both phases, since the volume 
at hadronization is not unambiguously known. Therefore, we cannot 
convert the entropy density determined on the lattice into a total 
entropy. Furthermore, lattice calculations do not tell us the 
number of (quasi-)particles, because there is no lattice definition 
of particle density. Therefore, a more detailed comparison of the 
entropy content of the hadronic phase and the quark phase  
at hadronization remains as a problem for future investigations.  

We now return to the calculation of net charge fluctuations in
a bulk recombination scenario. The fluctuations of the net charge
$Q=\sum_i q_i n_i$ are given by
\begin{eqnarray}
\langle \delta Q^2 \rangle & \equiv & \langle Q^2 \rangle 
- \langle Q\rangle ^2 \nonumber \\
& = &  \sum_i (q_i)^2 \langle n_i \rangle 
+\sum_{i,k} c^{(2)}_{ik} \langle n_i\rangle \langle n_k \rangle  q_iq_k 
\label{delta_q1}
\end{eqnarray}
where
$c^{(2)}_{ik}$ are the normalized two-particle correlation functions:
\begin{eqnarray}
c^{(2)}_{ii} &=& \frac{\langle n_i(n_i-1)\rangle }
                 {\langle n_i\rangle ^2} -1 \quad (i=k) \, ;\\
c^{(2)}_{ik} &=& \frac{\langle n_in_k\rangle}
                 {\langle n_i\rangle \langle n_k\rangle }-1 
\nonumber \\
             &=& \frac{\langle ( n_i - \langle n_i \rangle )
                               ( n_k - \langle n_k \rangle ) \rangle }
                      {\langle n_i \rangle \langle n_k \rangle }
                 \quad (i\neq k). 
\label{correl}
\end{eqnarray}
The last expression in eq.~(\ref{correl}) shows that $c^{(2)}_{ik}$ 
is positive if there is a positive correlation between the quarks of 
flavors $i$ and $k$. In the absence of two-particle correlations, 
(\ref{delta_q1}) can be rewritten as:
\begin{equation}
\langle \delta Q^2 \rangle = \frac{4}{9} (N_u +  N_{\bar{u}})
     + \frac{1}{9} (N_d + N_{\bar{d}} + N_s + N_{\bar{s}}) \, ,
\label{delta_q2}
\end{equation}
where $N_i = \langle n_i \rangle$ denotes the average number of 
constituent quarks of flavor $i$.

Our strategy is now as follows: Knowing the number of final-state
charged hadrons within a given rapidity interval, $dN_{\rm ch}/dy$, we 
can extrapolate by means of the statistical hadronization model
\cite{stat_had,other_sm} to the thermal abundances of hadrons produced at
the critical temperature $T_c$. We can then determine the total
number and flavor distribution of valence quarks contained in these 
hadrons. Assuming valence quark recombination, using eq.\ (\ref{delta_q2}), 
and neglecting correlations, we can then calculate the expected 
net charge fluctuation at hadronization. The prediction for 
$\langle \delta Q^2 \rangle$ derived in this way can then be compared
with the measured value of this quantity.

We start from the measured charged multiplicity in central Au+Au 
collisions at $130 A\cdot$GeV mentioned above. Including all 
established meson and baryon resonances in the chemical freezeout,
we find that a total hadron number $dN_{\rm had}/dy = 507 \pm 35$ is 
needed at $T_c$ to generate this charged particle number, when all
sequential decays of unstable hadrons are taken into account. These 
507 hadrons contain a total of 1125 quarks and antiquarks.
Applying eq.~(\ref{delta_q1}), this gives
a prediction 
\begin{equation}
d\langle\delta Q^2\rangle_{\rm q}/dy = 286\pm 23 \, . 
\label{q2_quarks}
\end{equation}
The measured value of the hadronic net charge fluctuation is 
\cite{STAR_fluc}
\begin{eqnarray}
d\langle\delta Q^2\rangle_{\rm had}/dy  & = & 
  \frac{1}{4} D \times dN_{\rm ch}/dy \nonumber \\ 
& = & \frac{1}{4} \times (2.8\pm 0.05) \times 
 (526 \pm 2 \pm 36 ) \nonumber \\
& = & 368\pm 33 \, .
\label{q2_expt}
\end{eqnarray}
We note that the errors in eqs.~(\ref{q2_quarks}) and (\ref{q2_expt})
are strongly correlated because they derive, in part, from the same
uncertainty in the measured value of the charged particle multiplicity.
Clearly, the net charge fluctuations resulting from the recombination
of quarks derived in this way can explain only about 80\% of the
observed fluctuations. 

If, instead, we had included only the ground state nonet mesons and 
octet and decuplet baryons at chemical freezeout, the inferred 
number of hadrons at $T_c$ would have been larger, $dN_{\rm had}/dy = 
635 \pm 43$, because fewer decays contribute to the final multiplicity.
The number of recombining quarks would be accordingly larger, about
1377, yielding the increased prediction
\begin{equation}
d\langle\delta Q^2\rangle_{\rm q}/dy = 345\pm 29 \, , 
\label{q2_quarks_gs}
\end{equation}
which is close to the observed value eq.\ (\ref{q2_expt}).  

A third way of estimating the number of constituent quarks at the
moment of recombination is to start from the value $dS/dy \approx
4450$ derived by Pratt and Pal and use the calculated entropy per
hadron ($S/N = 5.15$) for a resonance gas at $T_c$ to obtain an
estimate for the number of hadrons at freezeout: $dN_{\rm had}/dy
= 864 \pm 78$. This value translates into an estimate for the
net charge fluctuations from quark recombination of
\begin{equation}
d\langle\delta Q^2\rangle_{\rm q}/dy = 487\pm 39 \, , 
\label{q2_quarks_pp}
\end{equation}
which lies significantly above the observed value. 
Considering the systematic uncertainties inherent in these estimates,
we may conclude that the observed net charge fluctuations
in Au+Au collisions at RHIC are compatible with
the mechanism of bulk hadronization via recombination of valence
quarks in the absence of significant net charge correlations among 
the quarks. However, a more detailed estimate on the possible production
of entropy in the hadronic phase would be desirable to better
constrain the analysis. Another source of uncertainty in
our analysis is the possibility of the net-charge fluctuations 
increasing modestly in the hadronic phase due to diffusion 
\cite{Shuryak:2000pd}. 

We now discuss a modified variant of the recombination process. Recently, 
quenched lattice QCD calculations have shown evidence for the existence 
of mesonic bound state correlations even above the critical temperature 
\cite{AsHa03,Bielefeld02,DaKaPeWe03,UmNoMa02}. Brown {\em et al.}\ argued
within an effective field theory that bound states of charmed quark 
mesons, light quark mesons and gluons exist above $T_c$ \cite{BrLeRhSh04}. 
These findings suggest that $qq$ and $q \bar{q}$ pairs may participate  
in hadronization mechanism as ``elementary'' constituents, just like 
individual quarks and antiquarks. In order to explore such a scenario,
we modified eq.~(\ref{delta_q1}) as follows:
\begin{eqnarray}
\langle \delta Q^2 \rangle 
 &=&   \sum_i (q_i)^2 \langle n_i \rangle 
 \\
 &&  + \sum_{ij} (q_i + q_j)^2 \langle n_{ij} \rangle 
     + \sum_{ij} (q_i - q_j)^2 \langle \bar{n}_{ij} \rangle , \nonumber  
\label{eq-delta_q1}
\end{eqnarray}
where $n_{ij}$ and $\bar{n}_{ij}$ are the number of $qq$ and $q\bar{q}$ 
pairs, respectively. For simplicity, we assume that the average number of 
$qq$ ($q\bar{q}$) pairs is proportional to the products of the individual
quark numbers: $\langle n_{ij} \rangle =\alpha N_{i}N_{j}$; 
$ \langle \bar{n}_{ij} \rangle =\beta N_{i}N_{\bar{j}}$,
where $\alpha$ and $\beta$ are the relative pairing weights. We have again 
neglected the correlation terms.  

The first term in eq.~(\ref{eq-delta_q1}) yields eq.~(\ref{delta_q2}). 
The second term, which denotes the contribution from diquarks, is given 
by  
\begin{eqnarray}
& \displaystyle \sum_{ij} & ( q_i + q_j )^2 \langle n_{ij} \rangle \nonumber \\
& = & \frac{16}{9} \alpha  N_u N_u 
    + \frac{4}{9} \alpha ( N_d N_d + N_s N_s + N_d N_s ) 
\nonumber \\ 
& + & \frac{1}{9} \alpha ( N_u N_d + N_u N_s ) 
    + ( q \rightarrow \bar{q}) ,
\label{Eq-qq}
\end{eqnarray}
while the third term, denoting the contribution from quark-antiquark
pairs, is: 
\begin{eqnarray}
& \displaystyle \sum_{ij} & ( q_i- q_j )^2
\langle \bar{n}_{ij} \rangle  \nonumber \\
& = & \beta \left ( N_d N_{\bar{u}} + N_s N_{\bar{u}} + N_u N_{\bar{d}}+ 
   N_u N_{\bar{s}} \right ).
\label{Eq-qqbar}
\end{eqnarray}
We can now constrain the parameters $\alpha$ and $\beta$ using experimental 
value of $\langle \delta Q^2 \rangle$ for the two cases discussed
above, counting only ground state hadrons or all known resonances
at chemical freezeout. The total number of quarks and antiquarks 
on the right-hand side of eq.~(\ref{eq-delta_q1}) is constrained
to be the same as that obtained from the statistical model.      
      
\begin{figure}
\label{Fig-qq-qqb}
\centerline{
\includegraphics[width=0.5\linewidth]{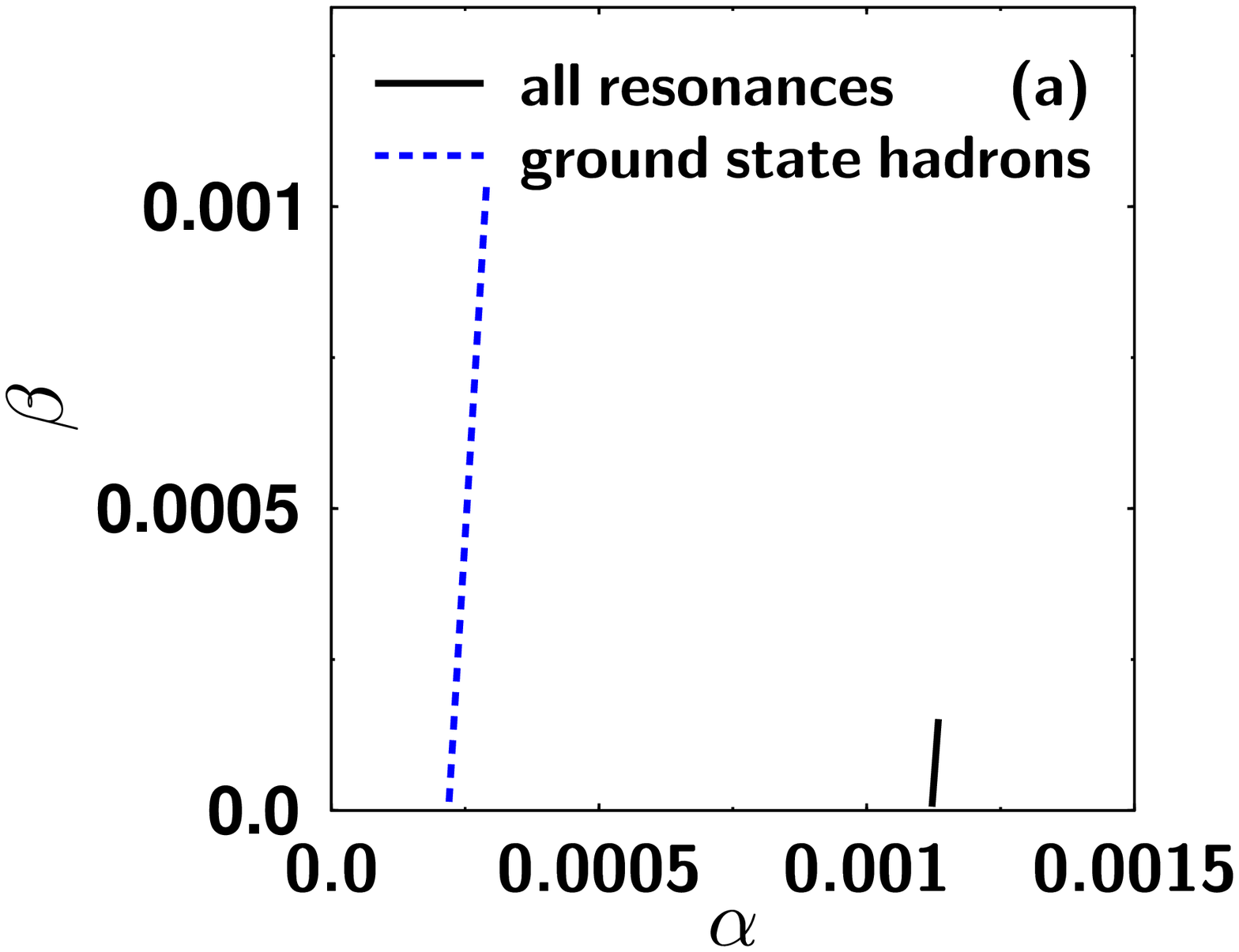}
\includegraphics[width=0.5\linewidth]{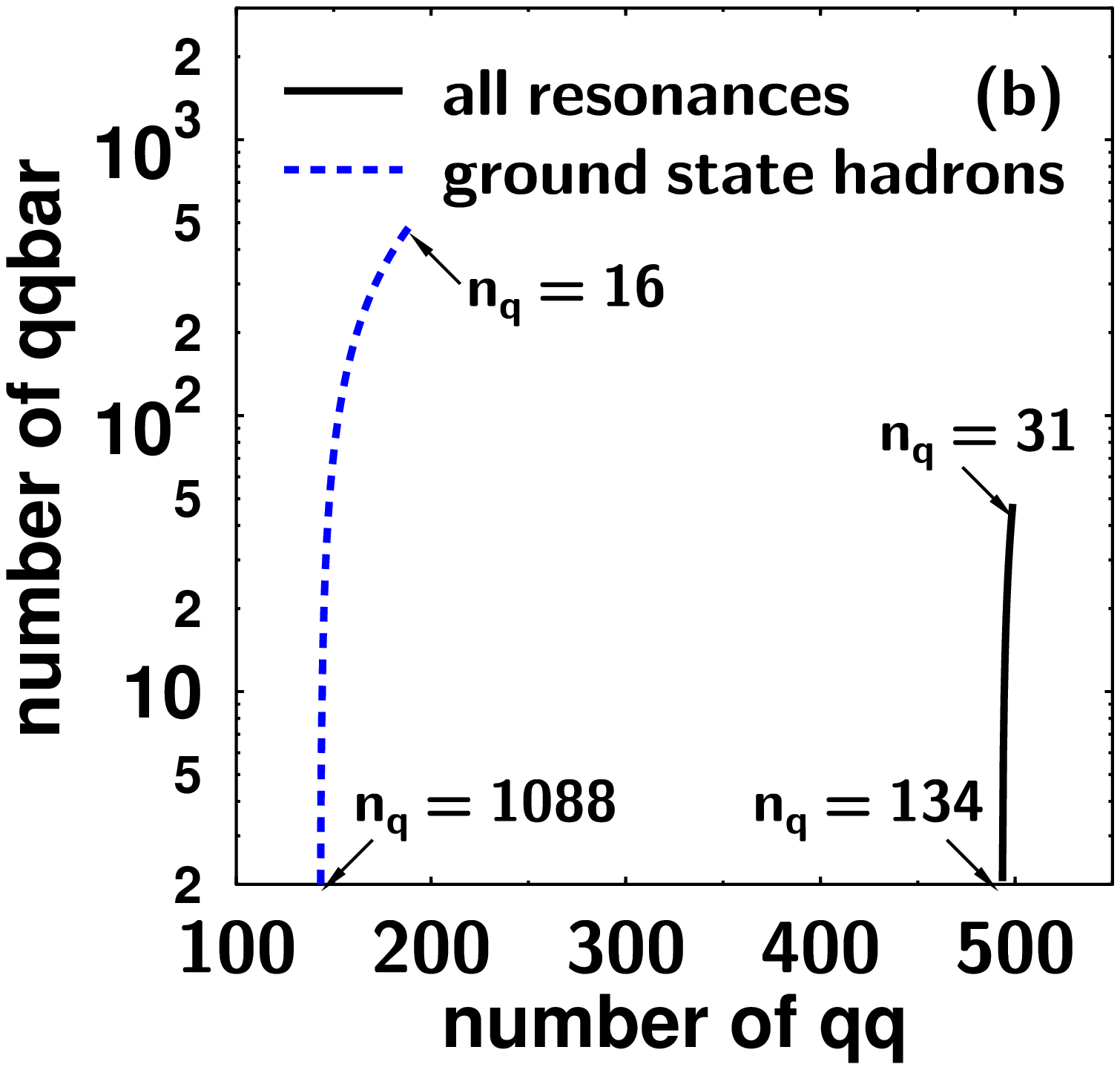}
}
\caption{
(a)
Relation between the weight of the contributions from $qq$ and 
$q\bar{q}$ pairs in the recombination process. The solid (dashed) 
line corresponds to a fixed total number of quarks and antiquarks 
of 1125 (1377).
(b)
Relation between the number of individual quarks (antiquarks)
and the number of diquarks or $q\bar{q}$ pairs participating in 
the recombination process. $n_q$ is the number of quarks and antiquarks 
contributing to the first term of eq.~(\ref{eq-delta_q1}).}
\end{figure}
 
Figure 1(a) shows the relation between the weights of 
$qq$ and $q\bar{q}$ pairs. This calculation is done in the 
simplest case ($N_u=N_{\bar{u}}=N_d=N_{\bar{d}}$, $N_s=N_{\bar{s}}$).   
The important point is the existence of the region 
where both $\alpha$ and $\beta$ are positive, which confirms
the possibility of a contribution from $qq$ and $q\bar{q}$ pairs
in the hadronization mechanism. We observe a similar tendency in 
both cases, counting ground state hadrons or all resonances. Diquark 
pairs are more favored than $q\bar{q}$ pairs in the sense that a
solution with $\beta=0$ is possible, but one with $\alpha=0$ is not. 
In fact, the value of $\beta$ is not well constrained by the charge 
fluctuations, because the difference between the charge of $qq$ and 
$q\bar{q}$ in eqs.\ (\ref{Eq-qq}) and (\ref{Eq-qqbar}). For 
example, in the simplest case, 
$ \sum_{ij} ( q_i + q_j )^2 \langle n_{ij} \rangle    
\sim \frac{42}{9}\alpha N^2_u$ and 
$\sum_{ij} ( q_i-\bar{q_j} )^2 \langle \bar{n}_{ij} \rangle 
 \sim 2 \beta N^2_u$, which implies that $qq$ pairs are favored by
a factor of about 2.  Perturbative QCD suggests that the
$q\bar{q}$ channel is more attractive than the $qq$ channel, 
implying $\alpha < \beta$ in eq.\ (\ref{eq-delta_q1}). 
In Fig.\ 1(a) the lower part of the dashed line 
and the whole solid line are forbidden if this condition is imposed. 
However, since the lattice results indicate that hadronization occurs 
via strong interactions between plasma quasi-particles, it is not
clear that this perturbative argument is applicable.

In order to clarify the relative numbers of diquarks, quark-antiquark
pairs, and individual (anti-)quarks participating in the recombination 
process, we plot the relation among $\sum_i N_i$ for quarks and 
antiquarks, $\sum_{ij}N_{ij}$ for diquarks, and $\sum_{ij}\bar{N}_{ij}$ 
for quark-antiquark pairs in Fig.~1(b). In the full 
resonance gas scenario, for example, we have $\sum_i N_i = 134$,
$\sum_{ij}N_{ij} = 493$, and $\sum_{ij}\bar{N}_{ij} = 2$, showing that 
$qq$ clustering dominates. The numbers of quarks and antiquarks decrease 
linearly as those of diquarks and quark-antiquark pairs increase along 
the solid and dashed lines, respectively. In the ground state hadrons only 
scenario, we find a region where recombination from individual quarks
and antiquarks is dominant. On the other hand, in the full resonance 
gas scenario, hadrons are predominantly created from diquark or
quark-antiquark clusters, which is difficult to reconcile with the 
elliptic flow data from RHIC \cite{STAR_reco,PHENIX_reco}, since that data
strongly suggests a constituent quark counting rule 
\cite{Vo02,Fr03,Gr03,Mo03}.
If we demand that the weight of individual quarks and    
antiquarks in the recombination process should be so large that 
the quark number scaling approximately survives, the full resonance 
gas scenario is disfavored.

In summary, 
we have investigated charged particle fluctuations at RHIC in the 
framework of the parton recombination model of hadronization
and find that within the present systematic uncertainties parton
recombination is compatible with the measured charged particle
fluctuations. 
We found that the behavior of the entropy density for an interacting 
deconfined system close to $T_c$ and the entropy per particle for 
the massive resonance gas support the recombination picture.    
Finally, we have investigated the possibility of bound state correlations
above $T_c$ and find them consistent with the parton recombination
approach as well, albeit constrained by the valence
quark number scaling observed in the data.

\begin{acknowledgments}
The authors thank T.~Renk for critical discussions and for
communicating the results of statistical model calculations.
They also thank K.~Kanaya for the original numerical
data for Tables I and II.
This work was supported in part by RIKEN, Brookhaven National 
Laboratory, the U.~S.~Department of Energy (grants 
DE-AC02-98CH10886, DE-FG02-96ER40945 and DE-FG02-03ER41239), 
the Japanese Ministry of Education (grant-in-aid 14540255), and 
the National Science Foundation (grant NSF-INT-03-35392). 
\end{acknowledgments}


\begin{thebibliography}{99}

\bibitem{Asakawa:2000wh}
M.~Asakawa, U.~W.~Heinz, and B.~M\"uller,
Phys.\ Rev.\ Lett.\  {\bf 85}, 2072 (2000). 

\bibitem{Jeon:2000wg}
S.~Jeon and V.~Koch,
Phys.\ Rev.\ Lett.\  {\bf 85}, 2076 (2000). 

\bibitem{Shuryak:2000pd}
E.~V.~Shuryak and M.~A.~Stephanov,
Phys.\ Rev.\ C {\bf 63}, 064903 (2001). 

\bibitem{Bleicher:2000ek}
M.~Bleicher, S.~Jeon, and V.~Koch,
Phys.\ Rev.\ C {\bf 62}, 061902 (2000). 

\bibitem{Mrowczynski:2001mm}
S.~Mrowczy\'nski,
Phys.\ Rev.\ C {\bf 66}, 024904 (2002). 

\bibitem{Zaranek:2001di}
J.~Zaranek,
Phys.\ Rev.\ C {\bf 66}, 024905 (2002). 

\bibitem{Nystrand:2003ed}
J.~Nystrand, E.~Stenlund, and H.~Tydesjo,
Phys.\ Rev.\ C {\bf 68}, 034902 (2003).


\bibitem{STAR_fluc}
J.~Adams {\it et al.}  [STAR Collaboration],
Phys.\ Rev.\ C {\bf 68}, 044905 (2003); 
C.~A.~Pruneau  [STAR Collaboration],
[arXiv:nucl-ex/0304021];
G.~D.~Westfall  [STAR collaboration],
J.\ Phys.\ G {\bf 30}, S1389 (2004). 

\bibitem{PHENIX_fluc}
K.~Adcox {\it et al.}  [PHENIX Collaboration],
Phys.\ Rev.\ Lett.\  {\bf 89}, 082301 (2002);
J.~Nystrand  [PHENIX Collaboration],
Nucl.\ Phys.\ A {\bf 715}, 603 (2003). 

\bibitem{Sako:2004pw}
H.~Sako and H.~Appelsh\"auser  [CERES/NA45 Collaboration],
J.\ Phys.\ G {\bf 30}, S1371 (2004). 

\bibitem{Alt:2004ir}
C.~Alt {\it et al.}  [NA49 Collaboration],
[arXiv:nucl-ex/0406013].


\bibitem{Bialas:2002hm}
A.~Bia{\l}as,
Phys.\ Lett.\ B {\bf 532}, 249 (2002). 

\bibitem{Biro:1994mp}
T.~S.~Bir\'o, P.~L\'evai, and J.~Zim\'anyi,
Phys.\ Lett.\ B {\bf 347}, 6 (1995);
J.~Zimanyi, P.~Levai, and T.~S.~Biro,
Heavy Ion Phys.\  {\bf 17}, 205 (2003).

\bibitem{Vo02}
S.~A.~Voloshin,
Nucl.\ Phys.\ A {\bf 715}, 379 (2003).

\bibitem{Fr03}
R.~J.~Fries, B.~M\"uller, C.~Nonaka, and S.~A.~Bass,
Phys.\ Rev.\ Lett.\  {\bf 90}, 202303 (2003);
Phys.\ Rev.\ C {\bf 68}, 044902 (2003).

\bibitem{Gr03}
V.~Greco, C.~M.~Ko, and P.~L\'evai,
Phys.\ Rev.\ Lett.\  {\bf 90}, 202302 (2003);
Phys.\ Rev.\ C {\bf 68}, 034904 (2003).

\bibitem{Hw03}
R.~C.~Hwa and C.~B.~Yang,
Phys.\ Rev.\ C {\bf 67}, 034902 (2003);
Phys.\ Rev.\ C {\bf 67}, 064902 (2003).

\bibitem{Mo03}
D.~Molnar and S.~A.~Voloshin,
Phys.\ Rev.\ Lett.\  {\bf 91}, 092301 (2003).

\bibitem{Ko04}
P.~F.~Kolb, L.~W.~Chen, V.~Greco, and C.~M.~Ko,
Phys.\ Rev.\ C {\bf 69}, 051901 (2004).

\bibitem{PHENIX_reco}
S.~S.~Adler {\it et al.}  [PHENIX Collaboration],
Phys.\ Rev.\ Lett.\  {\bf 91}, 182301 (2003).

\bibitem{STAR_reco}
J.~Adams {\it et al.}  [STAR Collaboration],
Phys.\ Rev.\ Lett.\  {\bf 92}, 052302 (2004).

\bibitem{P+P}
S.~Pal and S.~Pratt,
Phys.\ Lett.\ B {\bf 578}, 310 (2004).

\bibitem{B+D}
S.~A.~Bass and A.~Dumitru,
Phys.\ Rev.\ C {\bf 61}, 064909 (2000).

\bibitem{CPPACS}
A.~Ali Khan {\it et al}. [CP-PACS Collaboration],
Phys.\ Rev.\ D {\bf 64}, 074510 (2001).

\bibitem{Bielefeld}
F.~Karsch, E.~Laermann, and A.~Peikert,
Phys.\ Lett.\ B {\bf 478}, 447 (2000).  


\bibitem{stat_had}
T.~Renk, Phys.\ Rev.\ C {\bf 68} 064901 (2003).  

\bibitem{other_sm}
P.~Braun-Munzinger, D.~Magestro, K.~Redlich, and J.~Stachel,
Phys.\ Lett.\ B {\bf 518}, 41 (2001);
F.~Becattini and U.~W.~Heinz,
Z.\ Phys.\ C {\bf 76}, 269 (1997)
[Erratum-ibid.\ C {\bf 76}, 578 (1997)];
J.~Cleymans and K.~Redlich,
Phys.\ Rev.\ C {\bf 60}, 054908 (1999);
J.~Rafelski and J.~Letessier,
Phys.\ Rev.\ Lett.\  {\bf 85}, 4695 (2000).


\bibitem{AsHa03}
M.~Asakawa and T.~Hatsuda, 
Phys.\ Rev.\ Lett. {\bf 92}, 012001 (2004). 

\bibitem{Bielefeld02}
F.~Karsch, E.~Laermann, P.~Petreczky, S.~Stickan, and I.~Wetzorke,
Phys.\ Lett.\ B {\bf 530}, 152 (2002).

\bibitem{DaKaPeWe03}
S.~Datta, F.~Karsch, P.~Petreczky, and I.~Wetzorke, 
Phys.\ Rev.\ D {\bf 69}, 094507 (2004).

\bibitem{UmNoMa02}
H.~Matsufuru, T.~Umeda, and K.~Nomura, 
arXiv:hep-lat/0401010.

\bibitem{BrLeRhSh04}
G.~E.~Brown, C.-H.~Lee, M.~Rho, and E.~Shuryak, 
J.\ Phys.\ G {\bf 30}, S1275 (2004). 

\end{thebibliography}
\end{document}